\documentclass[preprint,12pt]{elsarticle}
\usepackage[utf8]{inputenc}
\usepackage[T1]{fontenc}
\usepackage{csquotes}
\usepackage{amssymb}
\usepackage{url}
\usepackage{xcolor}
\usepackage{soul}

\begin{document}

\title{Generating and grading 34 Optimized Norm-Conserving Vanderbilt Pseudopotentials for Actinides and Super-Heavy Elements in the PseudoDojo}

\author[1,2,3]{Christian Tantardini\corref{cor1}}
\ead{christiantantardini@ymail.com}
\cortext[cor1]{Corresponding authors}
\cortext[contrib]{Ch.T., M.I. and M.G. contributed equally to this work}

\affiliation[1]{organization={Hylleraas center, Department of Chemistry, UiT The Arctic University of Norway},
            addressline={PO Box 6050 Langnes}, 
            city={Troms\o},
            postcode={N-9037}, 
            country={Norway}}

\affiliation[2]{organization={Department of Materials Science and Nonoengineering , Rice University},
            addressline={6100 Main St}, 
            city={Troms\o},
            postcode={77005},
            state = {Texas},
            country={USA}}

\affiliation[3]{organization={Institute of Solid State Chemistry and Mechanochemistry SB RAS},
            addressline={Kutateladze Street 18}, 
            city={Novosibirsk},
            postcode={630128},
            country={Russian Federation}}

\author[4,5,6]{Miroslav Ilia\v{s}}

\affiliation[4]{organization={Helmholtz-Institut Mainz},
            addressline={Staudingerweg 18}, 
            city={Mainz},
            postcode={55099},
            country={Germany}}

\affiliation[5]{organization={GSI Helmholtzzentrum f\"{u}r Schwerionenforschung GmbH},
            addressline={Planckstr. 1}, 
            city={Darmstadt},
            postcode={D-64291},
            country={Germany}}

\affiliation[6]{organization={Department of Chemistry, Faculty of Natural Sciences, Matej Bel University},
            addressline={Tajovského 40}, 
            city={Banská Bystrica},
            postcode={97401},
            country={Slovakia}}

\author[7]{Matteo Giantomassi\corref{cor1}}
\ead{matteo.giantomassi@uclouvain.be}

\affiliation[7]{organization={European Theoretical Spectroscopy Facility, Institute of Condensed Matter and Nanosciences, Universit\'{e} catholique de Louvain},
            addressline={Chemin des \'{e}toiles 8}, 
            city={Louvain-la-Neuve},
            postcode={B-1348},
            country={Belgium}}

\author[8]{Alexander G. Kvashnin}

\affiliation[8]{organization={Skolkovo Institute of
Science and Technology, Skolkovo Innovation Center},
            addressline={Bolshoy boulevard 30}, 
            city={Moscow},
            postcode={121205},
            country={Russian Federation}}

\author[5]{Valeria Pershina}

\author[7]{Xavier Gonze\corref{cor1}}
\ead{xavier.gonze@uclouvain.be}

\begin{abstract}
In the last decades, material discovery has been a very active research field driven by the need to find new materials for many different applications. 
This has also included materials with heavy elements, beyond the stable isotopes of lead, as
most actinides exhibit unique properties that make them useful in various applications.
Furthermore, new heavy elements beyond actinides, collectively referred to as 
super-heavy elements (SHEs), have been synthesized, filling previously empty space of Mendeleev periodic table.
Their chemical bonding behavior, of academic interest at present, would also benefit of state-of-the-art modeling approaches.

In particular, in order to perform first-principles calculations with planewave basis sets, one needs corresponding pseudopotentials.
In this work, we present a series of scalar- and fully-relativistic optimized norm-conserving Vanderbilt pseudopotentials (ONCVPs) for thirty-four actinides and super-heavy elements, for three different exchange-correlation functionals (PBE, PBEsol and LDA). 
The scalar-relativistic version of these ONCVPs is tested by comparing equations of states for crystals, obtained with \textsc{abinit} 9.6, with those obtained by all-electron zeroth-order regular approximation (ZORA) calculations,
without spin-orbit coupling, performed with the Amsterdam Modeling Suite \textsc{band} code. 
$\Delta$-Gauge and $\Delta_1$-Gauge indicators are used to validate these pseudopotentials.  
This work is a contribution to the PseudoDojo project, in which pseudopotentials for the whole periodic table are developed and systematically tested.
The pseudopotential files are available on the PseudoDojo web-interface pseudo-dojo.org in psp8 and UPF2 formats, both suitable for \textsc{abinit}, the latter being also suitable for Quantum ESPRESSO.
\end{abstract}

\flushbottom
\maketitle

\thispagestyle{empty}

\begin{figure}[ht]
\centering
\includegraphics[width=\linewidth]{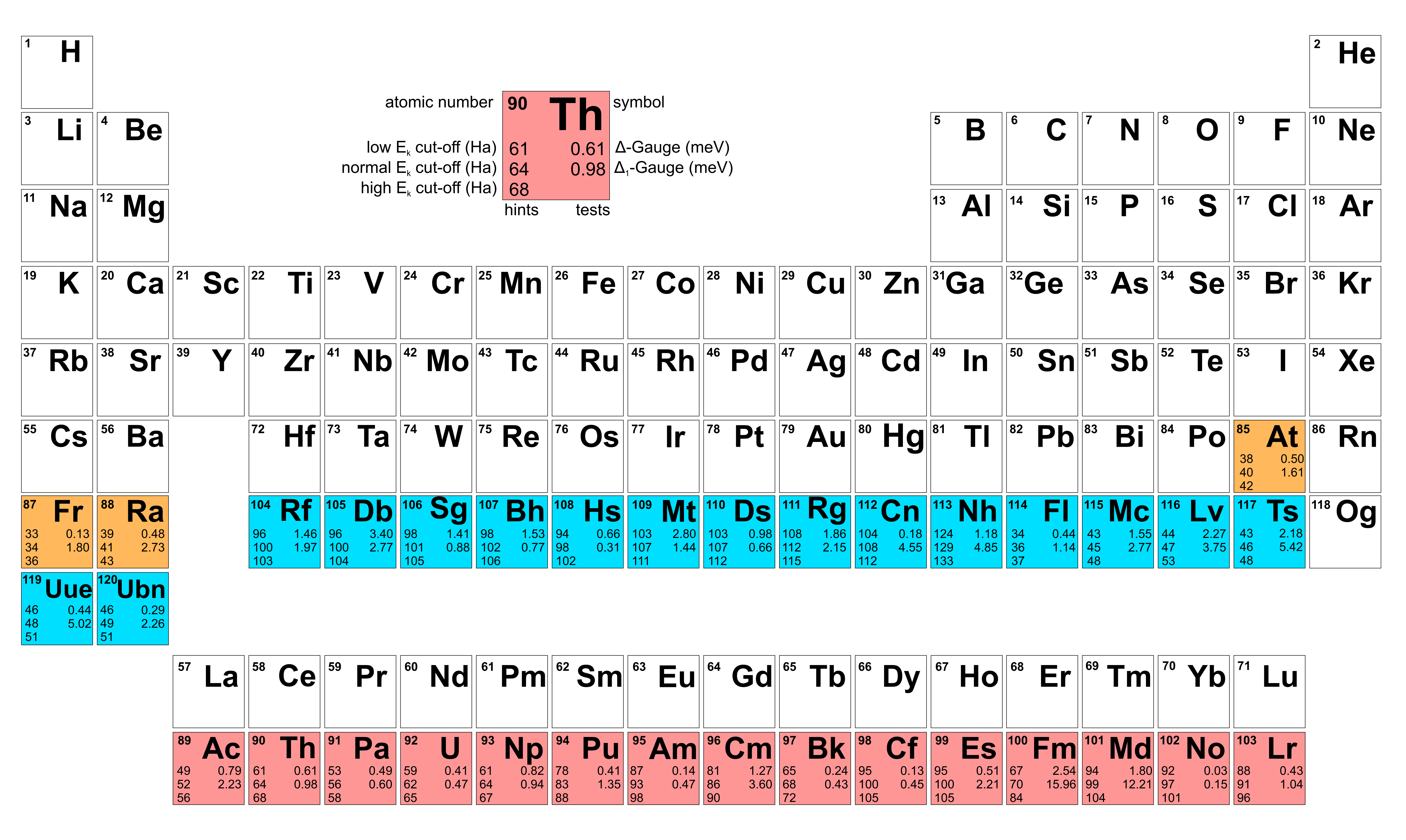}
\caption{Periodic table showing the 34 elements for which ONCVPs are developed in the present work.
Actinides are highlighted in red, SHEs in blue, and Fr, Ra and At in orange. For each element, the atomic number and symbol are mentioned in the upper part of the cell. In the lower left part of the cell, hints for kinetic energy cut-off are given, in Hartree. In the lower right part of the cell, the $\Delta$-Gauge and normalized $\Delta_1$-Gauge values are given, both in \textit{meV}.}
\label{fig:ptable}
\end{figure}

\section{Introduction}

In the last decades the search for new materials incorporating actinide elements has been a hot topic. 
They are useful in various applications like power engineering, medicine, industry, research, etc.
Some of them can form high-temperature superconducting hydrides under high pressure \cite{boeri_2022,struzhkin_superconductivity_2020, semenok_distribution_2020, trojan_high-temperature_2022}.
Actinides like thorium, uranium, and plutonium actively form oxides with a band gap close to that of gallium arsenide (GaAs), making them very efficient solar cells absorbers \cite{kumar_optical_2019,usov_uranium_2017}.
Furthermore, uranium and plutonium oxides are considered to be promising for high-density integrated circuits with higher breakdown voltages \cite{kumar_optical_2019,usov_uranium_2017, adamska_characterisation_2015} due to twice higher dielectric constant compared to GaAs (i.e., 14.1).
They can be used as bipolar junction transistor due to formation of the hole (i.e., \textit{p}-type) and electron (i.e., \textit{n}-type) polaron levels in the band gap in the presence of oxygen defects in the interstitial space, or in the case of oxygen vacancies \cite{Willardson_1958,Kruschwitz_2014,Khilla_1981,Ding_2020,Pena_2023,Li_2023,Dong_2023,Colmenares_1975,Larson_1980,Tang_2023,Shi_2023,Dinh_2023}.
 
Some actinides are widely present on earth, but most are scarcely abundant and all of them are radioactive, with a reduction of their scope for applications.
Notwithstanding this shortcoming, and considering more the academic interest, new opportunities appears thanks to the development and availability, even scarce, of SHEs, new heavier elements expanding the Mendeleev periodic table.
In the last decades new techniques were developed to synthesize SHEs with atomic number \textit{Z} larger than 103, approaching \textit{Z} equal to 120.
SHEs are found in \textit{s}-, \textit{p}- and \textit{d}-blocks of the Mendeleev periodic table. 
Unfortunately, SHEs have short lifetimes, which makes difficult the study of their chemistry \cite{Dullmann_2023,Das_2023,Zhu_2023,Holmbeck_2023,Turler_2015}, apart from theoretical studies.

Let us mention some of these theoretical studies, to illustrate the interesting chemistry (and physics) that might be found
with these SHEs. Indeed, the chemical behavior of two SHEs has been found to differ from what was previously expected from them: copernicium, Cn (\textit{Z}=112) \cite{Pershina_2016,Dmitriev_2017,Zaitsevskii_2010,Nazarewicz_2018,Aydinol_2019,Saamark_2023} and oganesson, Og (\textit{Z}=118) \cite{Matheson_2018,Nazarewicz_2018,De_2023,Pederson_2023,Ryzhkov_2023,Jerabek_2018}.

Copernicium has electronic configuration [Rn]$7s^{2} 5f^{14} 6d^{10}$.
This means that it belongs to the group 12 in modern IUPAC notation.
Other elements of group 12 are Zn, Cd, and Hg.
If the first two are seen to be solid, the last is liquid at normal conditions.
In Hg there are filled 4\textit{f} orbitals close to the nucleus which lead to reduced shielding of the nuclear charge on the valence shell.
This effect is called lanthanide contraction and is responsible for the high interaction between the 6\textit{s} orbitals and the nucleus.
The lanthanide contraction in Hg is responsible for its liquid state at normal conditions.
In Cn($Z$=112) the contraction affects the orbital energies, making energies of 5\textit{f}-orbitals closer to 6\textit{d}-orbitals.
This distinguishes Cn from the other elements of group 12  \cite{Pershina_2016,Dmitriev_2017,Zaitsevskii_2010,Nazarewicz_2018,Aydinol_2019,Saamark_2023,Nash_2005}.

The other is oganesson Og($Z$=118), the only super-heavy element that belongs to the noble gases family (i.e. group 18 of the periodic table according to IUPAC notation).
Og is expected to be a colorless, odorless, tasteless, non-flammable gas like all other noble gases.
However, the most important aspect of all noble gases is that they have a fully occupied outer electronic shell, which makes them reluctant to participate in chemical reactions under normal conditions.
In the last decades the field of high-pressure chemistry has shaken the established chemistry of noble gases, revealing new compounds with helium, sodium and others \cite{Dong_2017, Mercier_2023,Liu_2019,Shi_2020,Monserrat_2018,Zhang_2018,Gao_2020,Miao_2015,Gao_2019}, like the Na$_{2}$He stable crystal at pressure higher than 135 GPa \cite{Dong_2017,Mercier_2023,Liu_2019}.
Furthermore, new highly reactive compounds could be obtained from noble gases by photo-chemical reaction: fluorides like XeF$_{2}$ and KrF$_{2}$, or oxides like XeO$_{3}$ \cite{Liu_2017}.
In fact, Og($Z$=118) appears to form charged hydrides, fluorides and diatomic molecules with itself under various environmental conditions, making this element attractive for its high reactivity relative to the other noble gases. \cite{Matheson_2018,Nazarewicz_2018,De_2023,Pederson_2023,Ryzhkov_2023,Nash_2005}.

The chemical bonding of (most) actinides and SHEs with other elements has only been studied at the molecular level, albeit by all-electron (AE) approaches with complex relativistic post-Hartree-Fock methodologies and large local basis sets.
If post-Hartree-Fock approaches are fruitfully employed in gas phase they cannot be accurately employed for periodic systems. 
In fact, quantum chemical calculations for condensed matter usually rely on planewaves basis sets \cite{bloch_uber_1929}, where the core electrons of each atom are frozen and the divergence of Coulombic nuclear potential at the center of nucleus is avoided thanks to the use of a smooth pseudopotential.
Therefore, large periodic systems of actinides or SHEs are rarely studied with a planewave basis set due to the scarcity of pseudopotentials tailored for elements with \textit{Z} greater than 100.

Furthermore, actinides present a spatial extension of 5\textit{f}-orbitals comparable to 6\textit{d}-orbitals.
This means that the chemical bonding of actinides cannot be described if 5\textit{f}-electrons are frozen in the core, but they must be described by pseudowavefunctions in the valence shell, making more difficult the generation of pseudopotentials.
This is at variance with the one of 4\textit{f} electrons of the lanthanides, that for some materials can be frozen within the core of pseudopotential due to their smaller extension with little spatial overlap with the 5\textit{d}-electrons.
Thus, the developed pseudopotentials for actinides and SHEs should be characterized by numerous semicore states in the valence shell in order to increase the transferability of pseudopotential.
Such a warning applies to all pseudopotential formalisms, including the Projector-Augmented Wave (PAW) method, all giving uncontrolled error in the presence of overlap between pseudospheres.\cite{PAW}
As an example, in presence of overlap between the PAW pseudospheres, the computed electric field gradients might
have an error on the order of $10^{-1}$ \textit{MHz}, with unreliable sign \cite{ZWANZIGER201614}.
ONCVPs with semi-core states and small pseudocore radii can avoid such problems.
Moreover, ONCVPs open the possibility to use complex approaches as \textit{\textit{GW}} approximation \cite{Hedin_1965}, Bethe–Salpeter equation \cite{BSE} and electron-phonon coupling \cite{EPH1} for advanced studies of solid-state crystals.

\begin{table}
    \centering
    \begin{tabular}{c|c|c|c|c|c|c} \hline
Atom &  Z  & $V_{0}^{AE}$ / \AA$^{3}$  & $V_{0}^{PW}$ / \AA$^{3}$  & $V_{0}^{re}$ / \%   &$V_{0}^{PW}$ / \AA$^{3}$ & $V_{0}^{re}$ / \%   \\
     &     &          &  (\textsc{abinit})   & (\textsc{abinit})   & (\textsc{vasp}) & (\textsc{vasp})  \\ \hline
 At  &  85 &  39.048  &  38.922   &  -0.323  &  -  & - \\   
 Fr  &  87 & 117.023  & 116.730   &  -0.250  &  -  & - \\
 Ra  &  88 &  72.205  &  71.908   &  -0.411  &  -  & - \\
 Ac  &  89 &  45.686  &  45.569   &  -0.256  & 45.474  & -0.463  \\
 Th  &  90 &  32.309  &  32.296   &  -0.040  & 32.065  & -0.756  \\
 Pa  &  91 &  25.318  &  25.370   &   0.205  & 26.688  & 5.412   \\
  U  &  92 &  21.777  &  21.791   &   0.064  & 21.809  & 0.146   \\
 Np  &  93 &  19.363  &  19.363   &   0.000  & 19.356  & -0.035  \\
 Pu  &  94 &  28.037  &  27.986   &  -0.182  & 26.703  & -4.757  \\
 Am  &  95 &  34.634  &  34.623   &  -0.032  & 32.673  & -5.662  \\
 Cm  &  96 &  30.818  &  31.001   &   0.594  & 30.567  & -0.814  \\
 Bk  &  97 &  18.282  &  18.248   &  -0.186  &  - & - \\
 Cf  &  98 &  24.905  &  24.904   &  -0.004  & 24.048  &  -3.442  \\
 Es  &  99 &  23.022  &  23.078   &   0.243  & -  & - \\
 Fm  & 100 &  27.631  &  28.329   &   2.526  & -  & - \\
 Md  & 101 &  33.639  &  34.290   &   1.935  & -  & - \\
 No  & 102 &  39.535  &  39.532   &  -0.008  & -  & - \\
 Lr  & 103 &  30.064  &  30.023   &  -0.136  & -  & - \\
 Rf  & 104 &  24.635  &  24.724   &   0.361  & -  & - \\
 Db  & 105 &  21.282  &  21.373   &   0.428  & -  & - \\
 Sg  & 106 &  18.947  &  18.935   &  -0.063  & -  & - \\
 Bh  & 107 &  17.479  &  17.458   &  -0.120  & -  & - \\
 Hs  & 108 &  16.706  &  16.694   &  -0.072  & -  & - \\
 Mt  & 109 &  16.879  &  16.838   &  -0.243  & -  & - \\
 Ds  & 110 &  18.145  &  18.124   &  -0.116  & -  & - \\
 Rg  & 111 &  21.075  &  21.132   &   0.270  & -  & - \\
 Cn  & 112 &  43.124  &  42.738   &  -0.895  & 42.839  & -0.661  \\
 Nh  & 113 &  36.746  &  37.065   &   0.868  & 36.536  & -0.571  \\
 Fl  & 114 &  37.275  &  37.223   &  -0.140  & 37.444  &  0.453  \\
 Mc  & 115 &  37.570  &  37.414   &  -0.415  & 37.686  &  0.309  \\
 Lv  & 116 &  38.033  &  37.797   &  -0.621  & 38.433  &  1.051  \\
 Ts  & 117 &  43.312  &  42.876   &  -1.007  & 43.605  &  0.677  \\
Uue  & 119 & 100.652  &  99.874   &  -0.773  & -  & - \\
Ubn  & 120 &  78.470  &  78.543   &   0.093  & -  & - \\ \hline
    \end{tabular}
    \caption{Equilibrium volume $V_0$ of the Birch-Murnaghan\cite{Birch_1947} equation of state, from AE calculations (\textsc{band} code) and PW calculations using pseudopotentials from the present work - scalar-relativistic case (\textsc{abinit} code). Data from
     \textsc{vasp} are also provided, for the available PAW atomic datasets, see Methods section. The relative errors (re) with respect to the AE reference are also shown.}
    \label{tab:v-table}
\end{table}

\begin{table}
    \centering
    \begin{tabular}{c|c|c|c|c|c|c} \hline
Atom  & Z    &  $B_{0}^{AE}$ / GPa  & $B_{0}^{PW}$ / GPa  & $B_{0}^{re}$ / \%  & $B_{0}^{PW}$ / GPa  & $B_{0}^{re}$ / \%  \\
      &      &           & (\textsc{abinit})    &  (\textsc{abinit})  & (\textsc{vasp}) & (\textsc{vasp})  \\ \hline
 At   &  85  &   24.516  &   24.209 &  -1.252   &  - & -  \\   
 Fr   &  87  &    1.942  &    1.922 &  -1.030   &  - & -  \\
 Ra   &  88  &    7.022  &    7.281 &   3.688   &  - & -  \\
 Ac   &  89  &   23.746  &   23.207 &  -2.270   &  16.817   & -29.179 \\
 Th   &  90  &   55.211  &   57.883 &   4.840   &  58.109   & 5.249 \\
 Pa   &  91  &   95.267  &   95.144 &  -0.129   &  10.474   & -89.005 \\
  U   &  92  &  116.235  &  116.464 &   0.197   &  115.745  & -0.421 \\
 Np   &  93  &  136.602  &  135.551 &  -0.769   &  136.401  & -0.147 \\
 Pu   &  94  &   31.262  &   32.851 &   5.083   &  32.067   & 2.574 \\
 Am   &  95  &   24.992  &   25.550 &   2.233   &  16.797   & -32.792 \\
 Cm   &  96  &   31.597  &   34.101 &   7.925   &  33.759   & 6.842 \\
 Bk   &  97  &   88.695  &   92.014 &   3.742   & - & -  \\
 Cf   &  98  &   36.450  &   36.191 &  -0.711   &  39.807   &  9.210 \\
 Es   &  99  &   31.242  &   30.014 &  -3.931   & -  & - \\
 Fm   & 100  &   19.335  &   16.856 & -12.821   & -  & - \\
 Md   & 101  &   12.712  &   12.952 &   1.888   & -  & - \\
 No   & 102  &   13.887  &   13.644 &  -1.750   & -  & - \\
 Lr   & 103  &   39.098  &   41.384 &   5.847   & -  & - \\
 Rf   & 104  &   96.648  &   89.867 &  -7.016   & -  & - \\
 Db   & 105  &  170.686  &  171.898 &   0.710   & -  & - \\
 Sg   & 106  &  265.525  &  252.763 &  -4.806   & -  & - \\
 Bh   & 107  &  345.795  &  339.819 &  -1.728   & -  & - \\
 Hs   & 108  &  384.175  &  377.751 &  -1.672   & -  & - \\
 Mt   & 109  &  347.424  &  345.757 &  -0.480   & -  & - \\
 Ds   & 110  &  256.470  &  246.644 &  -3.831   & -  & - \\
 Rg   & 111  &  118.939  &  122.504 &   2.997   & -  & - \\
 Cn   & 112  &    2.728  &    2.847 &   4.362   &  2.704   &-0.878  \\
 Nh   & 113  &   19.103  &   19.727 &   3.267   & 19.371   & 1.403 \\
 Fl   & 114  &   32.242  &   31.178 &  -3.300   & 33.128   & 2.747 \\
 Mc   & 115  &   46.125  &   45.017 &  -2.402   & 44.754   & -2.973 \\
 Lv   & 116  &   47.214  &   48.112 &   1.902   & 45.542   & -3.541 \\
 Ts   & 117  &   26.811  &   28.105 &   4.826   & 27.450   &  2.384\\
Uue   & 119  &    2.574  &    2.612 &   1.476   & -  & - \\
Ubn   & 120  &    5.777  &    4.917 & -14.887   & -  & - \\ \hline
    \end{tabular}
    \caption{Bulk modulus $B_0$ of the Birch-Murnaghan\cite{Birch_1947} equation of state from AE calculations (\textsc{band} code) and PW calculations using pseudopotentials from the present work - scalar-relativistic case (\textsc{abinit} code). Data from
     \textsc{vasp} are also provided, for the available PAW atomic datasets, see Methods section. The relative errors (re) with respect to the AE reference are also shown.}
    \label{tab:b-table}
\end{table}

Thus, in the present work we continue the \verb|PseudoDojo| project (web interface at \url{pseudo-dojo.org}) started in 2018 \cite{vansetten_2018}, in which ONCVPs were proposed for 85 elements.
The \verb|PseudoDojo| project includes a Python framework for automatic generation and validation of pseudopotential properties. 
It consists of three different parts: 
(1) a database of reference results obtained with
AE (i.e. Gaussian orbital basis set) and planewave basis set codes; 
(2) a set of tools and graphical interfaces to facilitate the generation and initial validation of the pseudopotentials;
(3) a set of scripts to automate the execution of the various periodic structure tests with the \textsc{abinit} code. 

Here, we provide a set of fully-relativistic closed-shell ONCVPs for actinides and SHEs.
The precision of the developed ONCVPs is tested 
at the scalar-relativistic level by comparison with AE references, for
monoatomic face-centered cubic primitive cells,
based on the $\Delta$-Gauge \cite{Lejaeghere} and $\Delta_{1}$-Gauge\cite{Jollet_2014} descriptors.
The relative error for $V_{0}$ and $B_{0}$ between planewave and AE calculations is also shown.

\section{Methods}

\subsection{Generating the pseudopotentials} 
ONCVPs have been generated using the
version 4.0.1 of the ONCVPSP
package, available at 
\url{github.com/oncvpsp/oncvpsp}.
As mentioned in the previous section, design choices were such that a relatively large set of orbitals were
chosen as semi-core states, in order to
obtain good transferability and applicability of the generated 
pseudopotentials. Associated with
this design choice, the chosen cut-off radii have to be quite small.

Although the details of the ONCVP generation are given in Ref.~\cite{ONCVP1,ONCVP2}, some characteristics are worth to mention here.
To improve the transferability of pseudopotentials, we rely on Non-Linear Core Corrections (NLCC)\cite{Louie_1982}. 
NLCC removes the non-physical oscillations of the local potential $\hat{V}_{loc}$ (i.e., the potential coming from the projectors used to generate the pseudopotential), which must coincide with the AE potential $V_{AE}$ outside the pseudopotential cut-off radii $r_c$ \cite{ONCVP1,ONCVP2}.
Specifically, actinides are expected to have strong oscillation of the potential close to the nucleus, coming from the presence of 5\textit{f}-electrons in the valence shell.
We have used Teter NLCC\cite{Teter_1993} 
that allowed us to obtain smooth core charges with a consequent reduction in the kinetic energy cut-off required to obtain e.g. phonon convergence for the acoustic phonon branches.

Each channel corresponding to one specific $l$-state is characterized by its number of projectors, its cut-off radii, its number of Bessel functions and its cut-off wavevector of Bessel functions\cite{ONCVP1,ONCVP2}.
We provide such data in the Supporting Information, namely one table for each element, numbered ST.1 to ST.34.
The overall cut-off radius of the pseudopotential is the largest of the channel cut-off radii.

The number of projectors depends on the chosen valence electronic configuration and usually corresponds to the number of states for a specific angular momentum channel.
In some SHEs an unbounded \textit{f}-projector was added to improve the quality of pseudopotential.
The number of projectors as all other parameters chosen for the generation of a specific pseudopotential are given at the end of pseudopotential file for both formats psp8 and UPF2.

The number of projectors being defined, we have initially individually searched for each channel the cut-off radius that allows the pseudowavefunction to match the last maximum (or minimum) of the AE wavefunction, and we chose the necessary number of Bessel functions and their cut-off radii to allow the convergence of projectors so to obtain a good 
match for the logarithm derivative for $\ell$-state with respect to the energy between the pseudowavefunction and the AE wavefunction.
At the end, all channels are considered together for the final refinement of the pseudopotential. They are interconnected, because the variation of parameter of a channel inevitably affects the others.
The channel parameters are thus tuned in order to obtain both good $\Delta$-Gauge and low planewave kinetic energy cut-off for the total energy.
The parameters for the Teter model charge,\cite{ONCVP1,ONCVP2} which are called amplitude prefactor and scale prefactor, are initially chosen automatically by Nelder-Mead simplex algorithm \cite{Teter_1993} and subsequently manually adjusted to have a smoother model charge with consequent reduction of the needed kinetic energy cut-off and more stability of phonon calculations.

Optimization in this many-dimensional space (on the order of ten adjustable parameters, the prior choice of number of projectors per channel being understood) is non-trivial.
However, the task is critically split between separate channels first, for which only few parameters
need to be tuned concurrently.
Instead of trying to rely on 
non-linear optimization as 
attempted by Hansel and coworkers
\cite{Hansel2015}, we relied on 
human-driven optimization,  
 made efficient thanks to 
 rich and fast graphical representation tools, 
 available in the 
 \verb|PseudoDojo| project \url{github.com/PseudoDojo}. 
The pseudopotential files are available in the \verb|PseudoDojo| project (web interface at \url{pseudo-dojo.org}), see the ``Data availability statement'' section.

 \subsection{Spin-orbit coupling}

Two ONCVP versions are proposed for each element shown in the Mendeleev periodic table seen in the Fig.\ref{fig:ptable}: scalar-relativistic and 
fully-relativistic pseudopotentials. 
Because in pseudopotential calculations with plane waves, the deepest electronic states are not treated, 
the four-component fully-relativistic electronic wavefunction that describes electron (i.e. positive eigenvalues) and minority contributions from positrons (i.e. negative eigenvalues) in the fully-relativistic Dirac
approach can be reduced to two components, where only the electronic (majority) contributions are explicitly describe \cite{Koelling_1977,Takeda_1978}.
For such pseudopotential calculations, 
the scalar-relativistic case differs from the 
fully-relativistic one only by the presence of spin-orbit coupling in the latter.
Both versions of the developed ONCVPs are referred to as \enquote{stringent} on the basis of their accuracy, following the same notation used in the previous \verb|PseudoDojo| paper \cite{vansetten_2018}.
 
\subsection{Grading the pseudopotentials}
The $\Delta$-Gauge is a well-established method to
cross-compare two first-principles numerical implementations. 
It was formulated by Lejaeghere \textit{et al.} \cite{Lejaeghere}, 
who pointed that there is no absolute reference against which to compare results from planewaves implementations with different type of pseudopotentials or from different all-electron
implementations, all differing in practice by 
various
numerical approximations, none being approximation-free.
The $\Delta$-Gauge quantifies the difference between two DFT-predicted equations of state $E(V)$ for some system (represented by the $i$ subscript) in the following way:

\begin{equation} 
\label{eq:delta}
    \Delta_{i}(a,b) = \sqrt{
    \frac{
    \int_{0.94V_{0,i}}^{1.06V_{0,i}} 
    \big ( E_{b,i}(V)-E_{a,i}(V) \big )^{2} \,dV
    }  
    {0.12V_{0,a,i}}
    },
\end{equation}

\indent where $\Delta_{i}(a,b)$ is the root-mean-square difference between the $E(V)$ of methods \textit{a} (i.e., the reference approach) and \textit{b}, over a ±6 \% interval around the equilibrium volume $V_{0,a,i}$ obtained with the reference approach (i.e. the \textit{a} method). 
Thus, having chosen one approach as the reference, the total energy is computed for the same volumic range $E(V)$ for both approaches and compared. 

However, in practice, the integral in this equation is evaluated through numerical
means, using energies from seven equally spaced volumes, from 94\% to 106\% of the reference $V_{0}$ by steps of 2 \%,
to provide a fit using the Birch-Murnaghan\cite{Birch_1947} equation
of state for each approach.
This was not made explicit in the original publication\cite{Lejaeghere}. 
The actual protocol followed
in the \verb|PseudoDojo| project
relies on scripts available in the ``Delta calculation package'', version 3.1, available at the end of the \url{molmod.ugent.be/deltacodesdft} Web page.
The script \verb|eosfit.py| allows one to compute the Birch-Murnaghan\cite{Birch_1947} equation of states and its parameters as equilibrium volume $V_{0}$, bulk modulus $B_{0}$ and the first derivative of the bulk modulus with respect to pressure $B'_{0}$.
These parameters are used by \verb|calcDelta.py| to solve numerically the integral within Eq.(\ref{eq:delta}) and finally compute the $\Delta$-Gauge.

Another validation parameter, called normalized $\Delta$-Gauge, denoted by the symbol $\Delta_1$-Gauge, \cite{Jollet_2014} is also presented,

\begin{equation}
    \Delta_{1}\textrm{-Gauge} = \frac{V_{ref}B_{ref}}{V_{AE}B_{AE}} \Delta\textrm{-Gauge}.
    \label{eq:Delta1}
\end{equation}

\indent $\Delta_{1}$-Gauge is a scaled value of $\Delta$-Gauge with respect to a reference material having a specific equilibrium volume $V_{ref}$ and bulk modulus $B_{ref}$ to allow comparison between all elements.
These reference values (i.e. $V_{ref}$ and $B_{ref}$), chosen from the of $V_{0}$ and $B_{0}$ for 71 elements, are 30 $Bohr^{3}$ and 100 $GPa$ respectively. They were fixed in the previous work by Jollet \textit{et al.} \cite{Jollet_2014}. 
Keeping this definition, that does not include actinides and SHEs nevertheless, makes sense since this is in any case just choosing a reference value for the volume and the bulk modulus.

The scalar-relativistic approximation  \cite{Koelling_1977,Takeda_1978} used 
in pseudopotential planewave calculations is comparable with the zero-order regular approximation \cite{Filatov_2003,Chang_1986,Heully_1986,Lenthe_1993,Van_1994} (ZORA) to the full relativistic Hamiltonian, however considered without spin-orbit coupling.
We have produced scalar-relativistic ONCVPs and validated their results with those of the ZORA AE calculations without spin-orbit coupling.
Hereafter, when we write ZORA, we refer to AE calculations done with the scalar-relativistic
approximation without spin-orbit coupling.
Neither in the pseudopotentials case nor in the AE case a so-called non-relativistic treatment is done. 

Nevertheless, fully-relativistic versions
of the pseudopotentials are delivered,
in addition to these validated scalar-relativistic pseudopotentials.
Indeed, spin-orbit coupling plays an important role for actinides and SHEs and it is the suggested version for any planewave calculation that involves these elements. The scalar-relativistic version is only presented for their validation.

The validation between AE and pseudopotential results for calculations including spin-orbit coupling has been considered as well. 
However, several considerations prevented us
for doing so.
First, this comparison would have introduced an additional source of difficulty for the comparison between pseudopotentials calculations and AE ones. 
Second, this was not the methodology followed for the \verb|PseudoDojo| and $\Delta$-Gauge projects, that relied on scalar-relativistic
calculations for both the AE and the pseudopotential (or PAW) cases.
Finally, the specific effect of spin-orbit coupling might not be best tested by examining
the total energy or an equation of
state.
At variance, Huhn and Blum\cite{SOC} have 
computed spin-orbit splittings
at selected points in the Brillouin Zone, and compared
two AE codes, FHI-aims and WIEN2K. 
One might as well consider to study the
magnetic anisotropy, specifically for magnetic materials.
We feel this interesting question might be worth 
a separate study.

\section{Computational Details}

To validate the newly generated pseudopotentials, crystals with a face-centered cubic primitive cell containing one atom per cell are considered. DFT calculations are performed with
the PBE functional,\cite{Perdew_1996} the most used XC functional for the validation of pseudopotentials for periodic structures.
Note however, that ONCVPs have 
been generated also with the PBEsol and LDA exchange-correlation functionals.

\subsection{All-electron calculations (AE)}
ZORA AE calculations \cite{Filatov_2003,Chang_1986,Heully_1986,Lenthe_1993,Van_1994} are performed with the \textsc{band} software application from the Amsterdam Modelling Suite \cite{PhysRevB.44.7888,BAND,Wiesenekker_1988,Wiesenekker_1991,Franchini_2013,Franchini_2014}.
The primitive cell is optimized using the Broyden-Fletcher-Goldfarb-Shanno (BFGS) optimization algorithm \cite{Broyden,Shanno,Goldfarb,Steihaug}.
Fermi-Dirac smearing with temperature equal to 0.001 Ha is used in all calculations. 
The Brillouin zone is sampled with a \mbox{15$\times$15$\times$15} k-point grid centered in $\Gamma$-point.
AE calculations are performed with local Gaussian basis set triple zeta plus double polarization basis set describing the virtual orbital space.
The convergence criterion of self-consistent total energy is 1.0 $\cdot$ 10$^{-6}$ Ha.

\subsection{Planewaves basis set calculations}
Planewaves basis set calculations with PBE\cite{Perdew_1996} DFT functional are performed with \textsc{abinit} \cite{Gonze2016,Gonze_2020}, for the validation of the ONCVPs, and with \textsc{vasp}\cite{VASP1,VASP2,VASP3,VASP4}, for comparison with prior existing PAW data, using the same parameters for the electronic smearing and BZ sampling as those used for \textsc{band}.
The kinetic energy cut-off for each element is determined individually for each developed ONCVP.
For PAW, one relies on the atomic dataset from 
Torres \textit{et al.} \cite{Torres_2020,Torres_2021}
for the actinides, and Trombach \textit{et al.}\cite{Trombach_2019} for the SHE \textit{p}-elements. No other PAW for SHEs had been made available to our knowledge.
We use the highest kinetic energy cut-offs suggested by \textsc{vasp} for actinides, and suggested by Trombach \textit{et al.} \cite{Trombach_2019} for SHEs.
The total energy convergence criterion cut-off is equal to 1.0 $\cdot$ 10$^{-6}$ Ha.

\section{Results and Discussion}

\begin{table}[ht]
    \centering
    \begin{tabular}{c|c|c|c} \hline
     Atom & Z  & $\epsilon_{s-orb}$ / \textit{eV} & $\epsilon_{p-orb}$ / \textit{eV} \\ \hline
 Ac &  89 &  208.22 & - \\
 Th &  90 &  242.59 & - \\
 Pa &  91 &  225.71 & - \\
  U &  92 &  225.40 & - \\
 Np &  93 &  225.03 & - \\
 Pu &  94 &  264.98 & - \\
 Am &  95 &  351.10 & - \\
 Cm &  96 &  266.93 & - \\
 Bk &  97 &  266.23 & - \\
 Cf &  98 &  265.48 & - \\
 Es &  99 &  264.68 & - \\
 Mc & 115 &  133.96 & - \\
 Lv & 116 &  134.80 & 165.53 \\
 Ts & 117 &  135.63 & -  \\
Ubn & 120 &  156.73 & -  \\ \hline
    \end{tabular}
    \caption{List of elements for which the pseudopotential exhibit high positive-energy ghost states.  
    $\epsilon$ gives the energy (\textit{eV}) at which the ghost appears in the corresponding $s$ or $p$ channels. 
    No ghosts are found at lower energies. The lowest ghost appearing below 130 $eV$, the safe energy region is sufficiently large for the reliability of electronic ground state calculations, but also electronic excitations of moderate energy, including optical properties.}
    \label{tab:ghost-table}
\end{table}

Following the methodology described in Sec. 2, we have generated pseudopotentials for the
actinides and SHEs, for both scalar-relativistic and
fully-relativistic cases, and for the PBE, PBEsol and LDA
exchange-correlation functionals.
For their availability, see the ``Data availability statement'' section.
The validation presented in what follows, is done for the
scalar-relativistic case, and using the PBE functional.

The equilibrium volume $V_0$ (see Tab.\ref{tab:v-table}) and the bulk modulus $B_0$ (see Tab.\ref{tab:b-table}) are compared from the two EOSs through the relative errors between the values predicted by ONCVP and AE data. 
For each element considered, the $\Delta$-Gauge parameter calculated with the high kinetic energy cut-off is shown in the Mendeleev periodic table, Fig.\ref{fig:ptable}.
As in the \verb|PseudoDojo| project, the \textit{low}, \textit{normal} and \textit{high} cut-off kinetic energies (Ha) for the
planewave basis set are presented in the Mendeleev periodic table (see Fig.\ref{fig:ptable}) to give guidance to the user.
The first one (\textit{low}) is used for a quick calculation or as a starting point for the convergence studies.
The second one (\textit{normal}) is used as a good guess for high-throughput calculations.
The third one is the cut-off beyond which no significant changes in the results should be observed.

The validity of the newly developed ONCVPs 
has also been checked by searching for 
additional highly-localized positive-energy states (see Tab.\ref{tab:ghost-table}), so-called ``ghost states''.\cite{Gonze1990,Gonze1991}
In some cases, ghost states at high energies are indeed observed.
When pseudopotentials are used to calculate properties that require an accurate description of the unoccupied region, i.e. optical properties or \textit{GW} calculations, ghost states must be avoided below 100 \textit{eV} above the Fermi level.
In many cases it was possible to remove ghost states by tuning the characteristics of the second projector. 
It has also been found that adding more semicore states improves the quality of the logarithmic derivative at high energies.
Note that the ghost states listed in Tab.\ref{tab:ghost-table} are observed at an energy high enough for not causing any problems.

\subsection{Elements with s highest (partly) filled shell}
The elements with s highest (partly) filled shell are Fr($Z$=87), Ra($Z$=88), Uue($Z$=119), and Ubn($Z$=120).
The two latter ones, Uue($Z$=119) and Ubn($Z$=120), are SHEs and are the last synthesized elements \cite{Manjunatha_2022,Manjunatha_2021}.
The mere specification of the \textit{s} valence shell with a pseudowavefunction is not sufficient to correctly describe the chemistry of these elements at high pressures \cite{Dong_2017,He_2018,Saleh_2016},
as \textit{d} semicore states are involved in chemical bonding \cite{Dong_2017,He_2018,Saleh_2016}.
Thus, valence configurations $5s^{2}5p^{6}5d^{10}6s^{2}6p^{6}7s^{1}$ for Fr($Z$=87) and $5s^{2}5p^{6}5d^{10}6s^{2}6p^{6}7s^{2}$ for Ra($Z$=88) have been used.
Similarly, valence configurations $6s^{2}6p^{6}6d^{10}7s^{2}7p^{6}8s^{1}$ for Uue($Z$=119) and $6s^{2}6p^{6}6d^{10}7s^{2}7p^{6}8s^{2}$ for Ubn($Z$=120) have been used.
In particular, an improvement of the $\Delta$-Gauge has been observed by introducing two projectors for the empty 6\textit{f} orbitals in Fr($Z$=87) and two projectors for the empty 7\textit{f} orbitals in Uue($Z$=119).

\subsection{Element with p highest (partly) filled shell}
At($Z$=85) was predicted by Niels Bohr as the $5^{th}$ halogen in the Mendeleev periodic table \cite{Kibler_2007}.
Dmitri Mendeleev in his work left an empty space on At($Z$=85) position.
Such element has been synthesized in laboratory by bombarding bismuth-209 with alpha particles \cite{Kibler_2007}, for the first time in 1940.
In recent decades, with the need to find new materials, At($Z$=85) has been found to be a promising element for the development of radiopharmaceuticals \cite{Guerard_2021,Wilbur_2013,Champion_2011,Vlsser_1983}.
No pseudopotential for At had been proposed in the 2018 set from the \verb|PseudoDojo| project, and this omission is removed in the present work, in order
to increase the possibility of new studies on its chemistry and possible compounds.
Here the valence shell configuration is $5s^{2}5p^{6}5d^{10}5f^{14}6s^{2}6p^{5}$, adding also two projectors for the empty 5\textit{f} orbitals in order to increase the transferability of the pseudopotential.

The other \textit{p}-elements considered in this work are the SHEs Nh($Z$=113) Fl($Z$=114) Mc($Z$=115), Lv($Z$=116) and Ts($Z$=117).
If the \textit{p}-elements of the $6^{th}$ series are subjected to the lanthanide contraction due to the presence of filled 4\textit{f}-orbitals which allow them to freeze in the nucleus as they are close to it, the SHEs of the $7^{th}$ series are characterized by filled 5\textit{f}-orbitals which are more spatially extended than the 4\textit{f}-orbitals.
So 5\textit{f} orbitals have been introduced into the valence shell, making it look like $5s^{2}5p^{6}5d^{10}5f^{14}6s^{2}6p^{6}6d^{10}7s^{2}7p^{1-5}$ for all of them.

Unfortunately we have not succeeded in developing a decent ONCVP for Og(118), sufficiently accurate ones being generated only with an important increase of the kinetic energy cut-off.

\subsection{Element with \textit{d} highest (partly) filled shell}
SHEs with \textit{d} highest (partly) filled shell, similarly to the SHEs with \textit{p} highest (partly) filled shell, are characterized by large filled 5\textit{f}-orbitals which require their inclusion in the valence shell. 
The series starts at Rf($Z$=104) and ends at Cn($Z$=112).
Their valence configuration is chosen to be \\
$5s^{2}5p^{6}5d^{10}5f^{14}6s^{2}6p^{6}6d^{1-10}7s^{2}$.
The 6\textit{d}-elements are characterized by $6d^{1-10}7s^{2}$ orbitals close in energy which are systematically involved in chemical bonding.
By including such extended semicore states we are confident of adequately describing the chemistry of such elements.

Triggered by the presence of ferromagnetic ordering for lighter elemental solids with \textit{d} highest (partly) filled shell, we have investigated the possibility to find a ferromagnetic ground state, for the FCC primitive cell, using AE calculations.
This was also done for  other SHEs and actinides.
None of SHEs showed such a magnetic behavior. This agrees with the relatively large spatial extension of partially filled 6\textit{d}-orbitals.
In the literature, we have not seen an investigation of the magnetic order for SHEs until now of their solid phase.

\begin{table}[ht]
    \centering
    \begin{tabular}{c|c|c|c} \hline
Atom & Z &   ONCVP Valence shell  & PAW Valence shell  \\    \hline
Ac &  89 & $5s^{2}5p^{6}5d^{10}6s^{2}6p^{6}6d^{1}7s^{2}$                &      $(11 e^{-})$                  \\           
Th &  90 & $5s^{2}5p^{6}5d^{10}5f^{0.10}6s^{2}6p^{6}6d^{1.90}7s^{2}$    &      $(12 e^{-})$                  \\             
Pa &  91 & $5s^{2}5p^{6}5d^{10}5f^{2}6s^{2}6p^{6}6d^{1}7s^{2}$          &      $(13 e^{-})$                  \\             
 U &  92 & $5s^{2}5p^{6}5d^{10}5f^{3}6s^{2}6p^{6}6d^{1}7s^{2}$          &      $(14 e^{-})$                  \\                  
Np &  93 & $5s^{2}5p^{6}5d^{10}5f^{4}6s^{2}6p^{6}6d^{1}7s^{2}$          &      $(15 e^{-})$                  \\                 
Pu &  94 & $5s^{2}5p^{6}5d^{10}5f^{5.90}6s^{2}6p^{6}6d^{0.10}7s^{2}$    &      $(16 e^{-})$                  \\                    
Am &  95 & $5s^{2}5p^{6}5d^{10}5f^{6.90}6s^{2}6p^{6}6d^{0.10}7s^{2}$    &     $5f^{7}6s^{2}6p^{6}7s^{2}$         \\            
Cm &  96 & $5s^{2}5p^{6}5d^{10}5f^{7}6s^{2}6p^{6}6d^{1}7s^{2}$          &     $5f^{6}6s^{2}6p^{6}6d^{2}7s^{2}$   \\         
Bk &  97 & $5s^{2}5p^{6}5d^{10}5f^{8.90}6s^{2}6p^{6}6d^{0.10}7s^{2}$    &      -                                 \\               
Cf &  98 & $5s^{2}5p^{6}5d^{10}5f^{9.90}6s^{2}6p^{6}6d^{0.10}7s^{2}$    &     $5f^{8}6s^{2}6p^{6}6d^{2}7s^{2}$   \\               
Es &  99 & $5s^{2}5p^{6}5d^{10}5f^{10.90}6s^{2}6p^{6}6d^{0.10}7s^{2}$   &      -                                 \\                
Fm & 100 & $5s^{2}5p^{6}5d^{10}5f^{12}6s^{2}6p^{6}7s^{2}$               &      -                                 \\               
Md & 101 & $5s^{2}5p^{6}5d^{10}5f^{13}6s^{2}6p^{6}7s^{2}$               &      -                                 \\   
No & 102 & $5s^{2}5p^{6}5d^{10}5f^{13.90}6s^{2}6p^{6}6d^{0.10}7s^{2}$   &      -                                 \\
Lr & 103 & $5s^{2}5p^{6}5d^{10}5f^{14}6s^{2}6p^{6}6d^{1}7s^{2}$         &      -                                 \\ \hline
    \end{tabular}
    \caption{For actinides, valence shell (including semi-core states) of present work ONCVPs  and valence shell of PAW \textsc{vasp} atomic data. The latter did not contain the details of the valence shell for elements from Ac to Pu. In this case we have indicated only the declared number of electrons. One might hypothesize that the $6s$, $6p$ and $7s$ shells of these elements are complete, leaving the incomplete $5f$ shell for the remaining electrons.}
    \label{tab:act-val}
\end{table}

\subsection{Actinides}
Actinides, with $Z$ from 89 to 103, are characterized by the presence of 5\textit{f}-electrons that play a major role in the chemical bonding. 
This differs from their lanthanide counterparts, the 5\textit{f}-orbitals having a large spatial extension, unlike lanthanide 4\textit{f}-orbitals.
Indeed, 4\textit{f}-orbitals are close to the nucleus and can often be frozen within the core, for materials in which the lanthanide has an oxidation state of III. \cite{vansetten_2018}

In the actinide series additional 5\textit{f}-orbitals are not systematically filled with increasing $Z$.
In fact we had to tailor the ONCVPs for the specific known valence shell of each actinide (see Tab.\ref{tab:act-val})

The presence of partly filled 5\textit{f}-orbitals often induces magnetism of actinide-based materials.
As a consequence, the magnetism of the FCC elemental crystal has been checked for each actinide, by performing ferromagnetic AE ZORA calculations \cite{Filatov_2003,Chang_1986,Heully_1986,Lenthe_1993,Van_1994}, see Tab.\ref{tab:act-mag}. In this table, it is seen that ten actinides exhibit magnetic behavior within the FCC primitive cell. 
This brought us to consider whether pseudopotential and AE EOS should be compared for the non-magnetic case or, instead, for the magnetic case.

\begin{table}[ht]
    \centering
    \begin{tabular}{c|c|c} \hline
Atom & Z & Hirshfeld mag. / $\mu B$ \\   \hline 
Ac &  89 &   -  \\              
Th &  90 &   -  \\                
Pa &  91 &   -  \\                
 U &  92 & -0.892 \\                   
Np &  93 & -2.695 \\                  
Pu &  94 & -5.785 \\                     
Am &  95 & -7.321 \\             
Cm &  96 & -7.026 \\          
Bk &  97 & -5.459 \\                
Cf &  98 & -4.164 \\                
Es &  99 & -2.906 \\                 
Fm & 100 & -1.643 \\                
Md & 101 & -0.419 \\    
No & 102 &   - \\   
Lr & 103 &   -  \\ \hline
    \end{tabular}
    \caption{Magnetic moment of actinides within FCC phase, in $\mu B$, calculated with ZORA AE calculations using atomic Hirshfeld partition.}
    \label{tab:act-mag}
\end{table}

Several concerns are present.
The $\Delta$-Gauge project relies only on EOS for the non-magnetic case, even for transition metals. 
This would favour relying on such non-magnetic EOS for actinides as well, for consistency reasons.
However, the $\Delta$-Gauge project did not include rare-earth elements with \textit{f} highest (partly) filled shell, for which the magnetism is particularly strong.
One can wonder whether studying EOS of non-magnetic phases of these materials is relevant at all.
The choice to restrict to non-magnetic EOS might not be completely pertinent not only for rare-earth elements, but also for the present actinides.

This being said, the specific choice of the FCC phase might anyhow bias the presence of magnetism.
On a more practical point of view, it was also observed that the determination of the AE EOS for several actinides is not straightforward due to the presence of two electronic phases in the magnetic case: 
the system jumps from one to the other, in the volume region where the energy is minimal, making difficult the 
application of the usual protocol to find the $\Delta$ factor.

Finally, we decided to stick to the non-magnetic treatment with most materials, with the exception of Pu($Z$=94), Am($Z$=95), Cm($Z$=96) and Cf($Z$=98). For these elements, only one stable electronic phase had been obtained in the magnetic case.
Ferromagnetic planewave and AE calculations for these gave a $\Delta$-Gauge on the order of 1 \textit{meV} or less, as presented in  Fig.\ref{fig:ptable}.

\subsection{ONCVP vs PAW}
Another methodology for pseudopotential generation is based on the projector augmented wave (PAW) formalism \cite{PAW}.
PAW pseudopotentials have atomic function projectors
and pseudoorbitals from which the all-electron (AE) wave function of an electronic state can be recovered, unlike with norm-conserving pseudopotentials.\cite{PAW}
Also, PAW pseudopotentials are included in the class of ultrasoft pseudopotentials, which in turn require lower kinetic energy cut-offs compared to more rigid norm-conserving pseudopotentials.
Although PAW faces challenges when being used beyond first-principles calculations, as described in the Introduction section, it is the most accurate
pseudization methodology and it is widely used for ground-state calculations.

ONCVP results have been compared with the PAW results obtained using the \textsc{vasp} code. \cite{VASP1,VASP2,VASP3,VASP4}. 
Actually, by the same token, these PAW results are compared with the current AE reference results.

There are only a few actinides for which PAW atomic data are available to perform \textsc{vasp} calculations. Moreover, the details of the electronic configuration of these is known only for Am ($Z$ = 95), Cm ($Z$ = 96), and Cf ($Z$ = 98), as shown in Tab.\ref{tab:act-val}.
As observed from the EOS generated for PAW \textsc{vasp}, the values of $V_{0}$ and $B_{0}$ are consistent with the results obtained from ONCVP. However, the latter showed much better agreement with AE results, as can be seen in Tab.\ref{tab:v-table} and Tab.\ref{tab:b-table}.
Other attempts to produce PAWs for actinides have been proposed in the literature, as suggested by Torres \textit{et al.} \cite{Torres_2020,Torres_2021}. 
These attempts employ equations of state (EOS) consistent with those used to produce PAWs with \textsc{vasp}. 

When it comes to SHEs, in the literature, only the PAW \textsc{vasp} atomic data for the \textit{p}-elements has been made available, by Trombach \textit{et al.}.\cite{Trombach_2019}
In this case, fitting the data to produce the EOS PAW \textsc{vasp} showed an unusual behavior for Cn($Z$=112) at the suggested kinetic energy cut-off. However, we found a smooth EOS when we increased it to 500 \textit{eV}.
Since we have not succeeded in finding a correct ONCVP capable of generating an EOS that could correctly fit the AE EOS, we have not examined the PAW \textsc{vasp} for the Og element ($Z$=118) any further. 
In contrast, for the other SHEs, the $V_{0}$ and $B_{0}$ obtained by PAW \textsc{vasp}\cite{Trombach_2019} agree reasonably well with the developed ONCVPs, as shown in Tab.\ref{tab:v-table} and Tab.\ref{tab:b-table}, giving confidence in the quality of both the ONCVP and PAW \textsc{vasp} atomic data sets.\cite{Trombach_2019}.

\section{Conclusion}
In conclusion, we have generated ONCVPs for thirty-four elements belonging to actinides and super-heavy elements, complementing the eighty-five ONCVPs of the original \verb|PseudoDojo| project \cite{vansetten_2018}.
The new ONCVPs were validated by considering the state-of-the-art descriptors $\Delta$-Gauge \cite{Lejaeghere} and $\Delta_{1}$-Gauge \cite{Jollet_2014}.
The developed scalar-relativistic ONCVPs for actinides show an average $\Delta$-Gauge equal to 0.73 \textit{meV} indicating a sufficiently high accuracy according to Ref.\cite{Lejaeghere_2016}.
For the SHEs we obtained an average $\Delta$-Gauge of 1.24 \textit{meV} showing a good accuracy according to Ref.\cite{Lejaeghere_2016}. 
The loss of accuracy from actinides to SHEs could be caused by the increase in relativistic effects, which might come from the differences between ZORA without spin-orbit coupling and scalar-relativistic pseudopotential approaches.
The developed ONCVPs are characterized by numerous \textit{s}- \textit{p}- \textit{d}- \textit{f}- semicore states, which make these pseudopotentials relatively hard (i.e., high kinetic energy cut-off), but suitable for approaches beyond ground state calculations such as the \textit{GW} approach.
The choice of small pseudopotential radii should allow these ONCVPs to be used successfully in high-pressure chemistry as well.
Developed ONCVPs provide a pathway for precisely predicting and studying material properties of elements containing actinides and SHEs, even under elevated pressure conditions. 

\section{Acknowledgments}
Ch.T. acknowledges support from the Research Council of Norway through its Centres of Excellence scheme (262695), through the FRIPRO grant ReMRChem (324590).
M. I. thanks the HIM for the financial support through the Visiting Scholar Agreement and for the support of the Slovak Research and Development Agency (APVV-20-0098). M. I. was partially supported by the grant of Ministry of Science and Higher Education of the Russian Federation, Nr. 075-10-2020-117.
The authors would like to acknowledge Pier Philipsen of Amsterdam Model Suite \textsc{band} for his support with the all-electron calculations and Davide Ceresoli of CNR of Milano for to the generation of UPF2 format of ONCVP for Quantum Espresso.

\section{Author contributions statement}
Ch. Tantardini and X. Gonze have initiated the project. 
Ch. Tantardini created the first version and several additional versions of all ONCVP pseudopotentials. 
The ONCVP pseudopotentials have been refined by M. Giantomassi based on these data to produce the current ones.
The ONCVP EOS were produced by M. Giantomassi using the specific \verb|PseudoDojo| python script that produced the hints and calculated the $\Delta$-Gauge and $\Delta_{1}$-Gauge.
M. Ilias has produced the AE EOS used as reference, supervised by V. Pershina. 
A.G. Kvashnin has produced the \textsc{vasp} PAW EOS pseudopotential results.
The manuscript has been written through contributions of all authors.
All authors have given approval to the final version of the manuscript.

\section{Data availability statement}
The pseudopotential files in their scalar relativistic and full-relativistic version for PBE, PBEsol and LDA are available in  \textsc{abinit} psp8 \cite{Gonze_2020,Gonze2016} and the Quantum ESPRESSO UPF2 formats at \cite{Giannozzi_2009,giannozzi_advanced_2017} \url{github.com/gmatteo/pseudos_ac_she}.
At the same link are showed the data coming from AE calculations with with the Amsterdam Modelling Suite \textsc{band} code as the input and output files for the generation of ONCVPs.
The output from \verb|oncvpsp| program contains the  data for visualization of $\ell$-states with their the logarithm derivative respect to the energy between the pseudowavefunctions and AE wavefunctions. It is also contained the energy cut-off for each pseudo-$\ell$-state. 
These data can be visualized with \verb|gnu-plot| script in \url{github.com/oncvpsp/oncvpsp}.  
Furthermore, we have included all data to generate the PBE equation of states for both scalar relativist ONCVP and ZORA AE as the subsequently files to compute $Delta$-Gauge and $Delta$-Gauge$_{1}$.

\bibliographystyle{elsarticle-num}
\bibliography{sample}

\end{document}